\documentclass[twocolumn,preprintnumbers,superscriptaddress,nofootinbib,aps,prd,floatfix]{revtex4-2}
\pdfoutput=1
\usepackage{enumerate}
\usepackage{gensymb}
\usepackage{amsmath,amssymb}
\usepackage{mathrsfs}
\usepackage{graphicx}
\usepackage{slashed}
\usepackage{xspace,slashed}
\usepackage[dvipsnames]{xcolor}
\usepackage[normalem]{ulem}
\usepackage{subfigure,orcidlink}
\usepackage[autostyle]{csquotes}
\usepackage{multirow,array}
\usepackage{float}
\usepackage{hyperref}
\hypersetup{colorlinks=true, citecolor=Blue, urlcolor=Blue, linkcolor=Blue}
\usepackage{tabularray}
\UseTblrLibrary{booktabs}
\usepackage[absolute,overlay]{textpos}

\newcommand{\orcidB}{\orcidlink{https://orcid.org/0000-0002-9107-5635}}
\newcommand{\orcidC}{\orcidlink{https://orcid.org/0000-0002-9675-0484}} 

\begin{document}

%%%%%%%%%%%%%%%%%%%
\title{Naturally small Dirac neutrino mass and $B-L$ dark matter}
%%%%%%%%%%%%%%%%%%%%

\author{Ernest Ma}
\affiliation{Department of Physics and Astronomy, University of California, Riverside, California 92521, USA.}

\author{Partha Kumar Paul\orcidB}
\email[Contact author:]{ph22resch11012@iith.ac.in}
\affiliation{Department of Physics, Indian Institute of Technology Hyderabad, Kandi, Sangareddy, Telangana-502285, India.}
	
\author{Narendra Sahu\orcidC}
\email[Contact author:]{nsahu@phy.iith.ac.in}
\affiliation{Department of Physics, Indian Institute of Technology Hyderabad, Kandi, Sangareddy, Telangana-502285, India.}    
	
%%%%%%%%%%%%%%%%%%%%%
\date{September 7, 2026}
\begin{abstract}
In the conventional gauged ${B-L}$ extension of the standard model, the $B-L$ charge of the singlet scalar $\chi$, responsible for the breaking of $U(1)_{B-L}$ symmetry, is taken to be 2 such that it can anchor type-I seesaw by giving Majorana masses to the right-handed neutrinos, $\nu_R$. In this paper, we consider instead the cases $\chi \sim 3$ or 4 under $B-L$, so that $\nu_R$ may not acquire any Majorana mass and neutrinos are Dirac fermions. We then consider a vector-like fermion $S$ with 2 units of $B-L$ charge, which becomes a good candidate for dark matter, either Dirac for $\chi \sim 3$ or Majorana for $\chi \sim 4$. In both cases, spontaneous $B-L$ breaking can induce a strong first-order phase transition, producing stochastic gravitational waves (GW) which can be tested at GW experiments. Moreover, the presence of light $\nu_R$s gives rise to an additional contribution to the effective number of relativistic degrees of freedom, $\Delta{N}_{\rm eff}$, providing complementary constraints from current and upcoming CMB observations.
\end{abstract}

%%%%%%%%%%%%%%%%%%%%%%%%%%%%%%%%%%%%%%%%%%%%%%%%%%%%%%%%%%%%%%%%%%%%%%	
\maketitle	
\preprint{\href{https://doi.org/10.1103/386s-sj3r}{10.1103/386s-sj3r}}
%%%%%%%%%%%%%%%%%%%%%%%%%%%%%%%%%%%%%%%%%%%%%%%%%%%%%%%%%%%%%%%%%%%%%%	
\section{Introduction}

To allow for nonzero neutrino masses, the standard model (SM) of quarks and leptons is routinely extended to include three right-handed neutrinos $\nu_R$.  As such, the model accommodates an additional well-known anomaly-free $U(1)$ gauge symmetry $B-L$.  The spontaneous breaking of $U(1)_{B-L}$ is usually assumed to come from a singlet scalar $\chi$ with two units of $B-L$ charge.  Hence $\nu_R$ acquires a large Majorana mass, and the type-I seesaw mechanism prevails \cite{Minkowski:1977sc, Gell-Mann:1979vob, Mohapatra:1979ia, Schechter:1980gr}.  However, there is a simple alternative.  If the $B-L$ charge of $\chi \sim 3$, then a residual global $U(1)$ lepton number remains conserved, and neutrinos are Dirac fermions.  This idea was first pointed out \cite{Ma:2013yga} in a different context, but it does not explain why these Dirac masses are so small.  To do this, we borrow another existing idea, as first pointed out years ago \cite{Ma:2000cc}.  Assume a $\mathcal{Z}_2$ discrete symmetry, under which $\nu_R$ is odd.  Add a second Higgs doublet $\eta=(\eta^+,\eta^0)$ which is also odd.  Require all dimension-4 terms in the Lagrangian to obey $\mathcal{Z}_2$, but break it softly by the quadratic term $\eta^\dagger \Phi$, where $\Phi=(\phi^+,\phi^0)$ is the SM Higgs doublet. With positive and large $\mu^2_\eta$ in the scalar potential \textit{i.e.} $\mu_\eta^2\eta^\dagger \eta$ term, the induced vacuum expectation value $\langle \eta^o \rangle$ is naturally small, thereby guaranteeing small Dirac neutrino masses.\footnote{See Refs.~\cite{Ma:2016mwh,Dasgupta:2019rmf} for examples of radiative Dirac neutrino mass generation and references therein.}

With the introduction of $B-L$ gauge symmetry, an interesting new scenario for dark matter (DM) emerges.  Suppose a singlet Dirac fermion $S(=S_L+S_R)$ is added with two units of $B-L$ charge.  It has an invariant mass, but does not interact with any SM particle except through the $B-L$ gauge boson. Note that the dimension-4 term $\chi \overline{S_L} \nu_R$ is forbidden by $\mathcal{Z}_2$. It is thus stable and is a good candidate for Dirac fermion DM.  In this paper, we consider this model, as well as a similar one where the $B-L$ charge of $\chi$ is 4 \cite{Heeck:2013rpa}. If the $B-L$ charge of $\chi$ is 4, then it can couple to a vector-like fermion $S$ as $\chi^\dagger \overline{S^c}S$, where the $B-L$ charge of $S$ is 2. Moreover, $S$ can have a Dirac mass term: $m\bar{S}S$. When $\chi$ gets a vacuum expectation value (vev), both $S_L$ and $S_R$ may acquire Majorana masses as well, resulting in two mixed eigenstates, the lighter of which is a candidate for Majorana fermion DM.

We mention in passing that in the canonical choice of $\chi \sim 2$ in the $B-L$ symmetric model, where the Majorana masses of light neutrinos are generated through type-I seesaw, the introduction of a Dirac fermion $S \sim 2$ under $B-L$ also works as dark matter. However, in this case, the dimension-5 $\chi^\dagger \chi^\dagger S_{L,R} S_{L,R}$ terms are admissible, making it only pseudo-Dirac. Other choices of $B-L$ charges for $\chi$ and $S$ are also possible for Dirac neutrino mass and dark matter, as can be easily worked out. 

%%%%%%%%%%%%%%%%%%%%%%%%%%%%%%%%%%%%%%%%%%%%%%%%%%%%%%%%%%%%%%%%%%%%%%
\section{Dirac neutrino mass and $\Delta{N}_{\rm eff}$}

The SM gauge group is extended with $U(1)_{B-L}$, which introduces non-zero chiral anomalies. These anomalies get automatically canceled once three right-handed neutrinos ($\nu_R$) are added with $U(1)_{B-L}$ charge -1. An additional $\mathcal{Z}_2$ symmetry is also introduced, under which $\nu_R$ is odd, to forbid the $\bar{L}\tilde{\Phi}\nu_R$ interaction. To forbid Majorana masses of RHNs, the $B-L$ charge of the singlet scalar $\chi$, responsible for breaking $U(1)_{B-L}$, is given a 3 or 4 charge. These two choices of $\chi$ charge lead to two different types of dark matter scenarios, as will be discussed later. To realize the Dirac neutrino mass, one $\mathcal{Z}_2$ odd scalar doublet $\eta=(\eta^+,\eta^0)$ is added with $B-L$ charge of 0. The $\mathcal{Z}_2$ symmetry is softly broken by $\mu_1^2\Phi^\dagger\eta$ interaction, which in turn results in naturally small Dirac neutrino mass. The particles and their charge assignments are shown in Table \ref{tab:tab1}.
\begin{table}[h]
\centering
\begin{tabular}{|c| c| c|} 
 \hline
 Particles & $U(1)_{B-L}$ & $\mathcal{Z}_2$\\
 \hline
$L$  & -1 &+ \\
 \hline
 $\Phi$  & 0 &+  \\
 \hline
 $\nu_R$  & -1 &- \\ 
 \hline
 $\eta$  & 0 &-  \\
 \hline
 $\chi$  & 3(4) &+  \\
 \hline
\end{tabular}
\caption{Particles and their charge assignments under $U(1)_{B-L}\times\mathcal{Z}_2$ symmetry.}\label{tab:tab1}
\end{table}
The relevant Yukawa Lagrangian can be written as 
\begin{equation}
    \mathcal{L}_{\rm Yukawa}=-y_R\bar{L}\tilde{\eta}\nu_R  + {\rm H.c.},
\end{equation}
The scalar Lagrangian is given as
\begin{equation}
    \mathcal{L}_{\rm scalar}=|D_\mu \Phi|^2+|D_\mu \eta|^2+|\mathcal{D}_\mu \chi|^2-V(\Phi,\eta,\chi)+ {\rm H.c.},
\end{equation}
where $D_\mu=\partial_\mu+i\frac{g}{2}\sigma_a W^a_\mu+i\frac{g^\prime}{2}B_\mu$ and $\mathcal{D}_\mu=\partial_\mu+i3(4)\textsl{g}_{\rm BL}(Z_{\rm BL})_\mu$, $\textsl{g}_{\rm BL}$ is the new gauge coupling. The scalar potential is given as
\begin{eqnarray}
    V(\Phi,\eta,\chi)&=&-\mu_h^2\Phi^\dagger \Phi+\lambda_h(\Phi^\dagger \Phi)^2+\mu_\eta^2\eta^\dagger\eta+\lambda_\eta(\eta^\dagger\eta)^2\nonumber\\&&-\mu_\chi^2\chi^*\chi+\lambda_\chi(\chi^*\chi)^2+\lambda_{1}(\Phi^\dagger \Phi)(\eta^\dagger\eta)\nonumber\\&&+\lambda_{2}(\Phi^\dagger \eta)(\eta^\dagger \Phi)+\frac{\lambda_{3}}{2}\left[(\Phi^\dagger \eta)^2+(\eta^\dagger \Phi)^2\right]\nonumber\\&&+\mu_1^2\Phi^\dagger\eta+\lambda_{h\chi}(\Phi^\dagger \Phi)(\chi^*\chi)\nonumber\\&&+\lambda_{\eta\chi}(\eta^\dagger\eta)(\chi^*\chi)+{\rm H.c.}
\end{eqnarray}

\begin{figure}[h]
    \centering
    \includegraphics[scale=0.35]{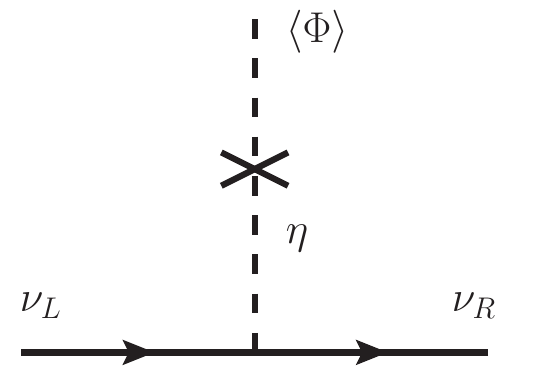}
    \caption{Tree-level Dirac neutrino mass.}
    \label{fig:neutrinomass}
\end{figure}

In the effective theory, the scalar fields can be parameterized as
\begin{eqnarray}
    \Phi=\begin{pmatrix}
        0\\
        \frac{h+v_h}{\sqrt{2}}
    \end{pmatrix},\eta=\begin{pmatrix}
        \eta^+\\
        \frac{\eta_R+v_\eta+i\eta_I}{\sqrt{2}}
    \end{pmatrix},\chi=\frac{\chi^\prime+v_\chi}{\sqrt{2}},
\end{eqnarray}
where $v_h,v_\eta,v_\chi$ are the vevs of $\Phi,\eta$, and $\chi$, respectively. We note that $\eta$ acquires an induced vev, $v_\eta\simeq \frac{\mu_1^2v_h}{\sqrt{2}m_{\eta_R}^2}$, which is naturally small by considering $\mu_1/m_{\eta_R}$ to be small. The $\eta_R$ and $\eta_I$ masses are related as $m_{\eta_R}^2-m_{\eta_I}^2=\lambda_3v_h^2$. Assuming $m_{\eta_R}\gg m_h,m_\chi$, we neglect the mixing between $h-\eta$ and $\chi-\eta$. The relevant scalar mixing for our purpose only exists between $h-\chi$. Mass matrix for the CP even states in the basis ($h~~\chi$) is
\begin{eqnarray}
    \mathcal{M}^2=\begin{pmatrix}
        2\lambda_hv_h^2 & \lambda_{h\chi}v_hv_\chi\\
        \lambda_{h\chi}v_hv_\chi &2\lambda_\chi v_\chi^2\\
    \end{pmatrix}.
\end{eqnarray}
The mixing angle is given as
\begin{eqnarray}
\tan2\gamma=\frac{\lambda_{h\chi}v_hv_\chi}{\lambda_\chi v_\chi^2-\lambda_h v_h^2}.
\end{eqnarray}
The Dirac neutrino mass can be realized at tree level as shown in Fig. \ref{fig:neutrinomass}.
Neutrino mass is given as
\begin{eqnarray}
    m_\nu= y_R v_\eta\simeq y_R\frac{1}{m_{\eta_R}^2}\mu_1^2\frac{v_h}{\sqrt{2}}.\label{eq:numass}
\end{eqnarray}
Assuming $m_\nu\sim0.05$ eV and $y_R\sim10^{-4}$, we get $\frac{\mu_1}{m_{\eta_R}}\sim5\times10^{-5}$.

In this scenario, the presence of light right-handed neutrinos $\nu_R$ can give rise to an additional contribution to the effective number of relativistic degrees of freedom (d.o.f), $\Delta{N}_{\rm eff}$. In the absence of these additional light d.o.f, the SM predicts a precise value for the effective number of relativistic species, namely $N_{\rm eff}^{\rm SM}=3.045$ \cite{Mangano:2005cc,Grohs:2015tfy,deSalas:2016ztq}. Depending on the value of $y_R$, both thermal and non-thermal production of $\nu_R$ can occur. For sufficiently large values of the coupling $y_R$, the $\nu_R$ states can be thermalized through processes such as $\nu_R\nu_L\rightarrow\nu_R\nu_L$, $\nu_R\eta^-\rightarrow\nu_R\eta^-$, $\nu_R\eta^+\rightarrow\nu_R\eta^+$, and $\nu_R\eta_{R/I}\rightarrow\nu_R\eta_{R/I}$. If the interaction rates of these processes fall below the Hubble expansion rate before the decoupling of the SM neutrinos, $\nu_R$ and $\nu_L$ subsequently evolve with different temperatures. The resulting population of relativistic $\nu_R$ then contributes to $\Delta{N}_{\rm eff}$, as given by \cite{Luo:2020sho}
\begin{eqnarray}
    \Delta{N}_{\rm eff}=N_{\rm eff}-N^{\rm SM}_{\rm eff}=N_{\nu_R}\left(\frac{g_{*s}(T^d_{\nu_L})}{g_{*s}(T^d_{\nu_R})}\right)^{\frac{4}{3}},
\end{eqnarray}
where the $T^d_{\nu_R}$ is the temperature at which $\nu_R$ decouples from the bath and $T^d_{\nu_L}$ is the SM neutrino decoupling temperature which is $\sim 1$ MeV, $N_{\nu_R}=3$ is the number of generations of $\nu_R$.
\begin{figure*}[tbh]
\centering
\includegraphics[scale=0.4]{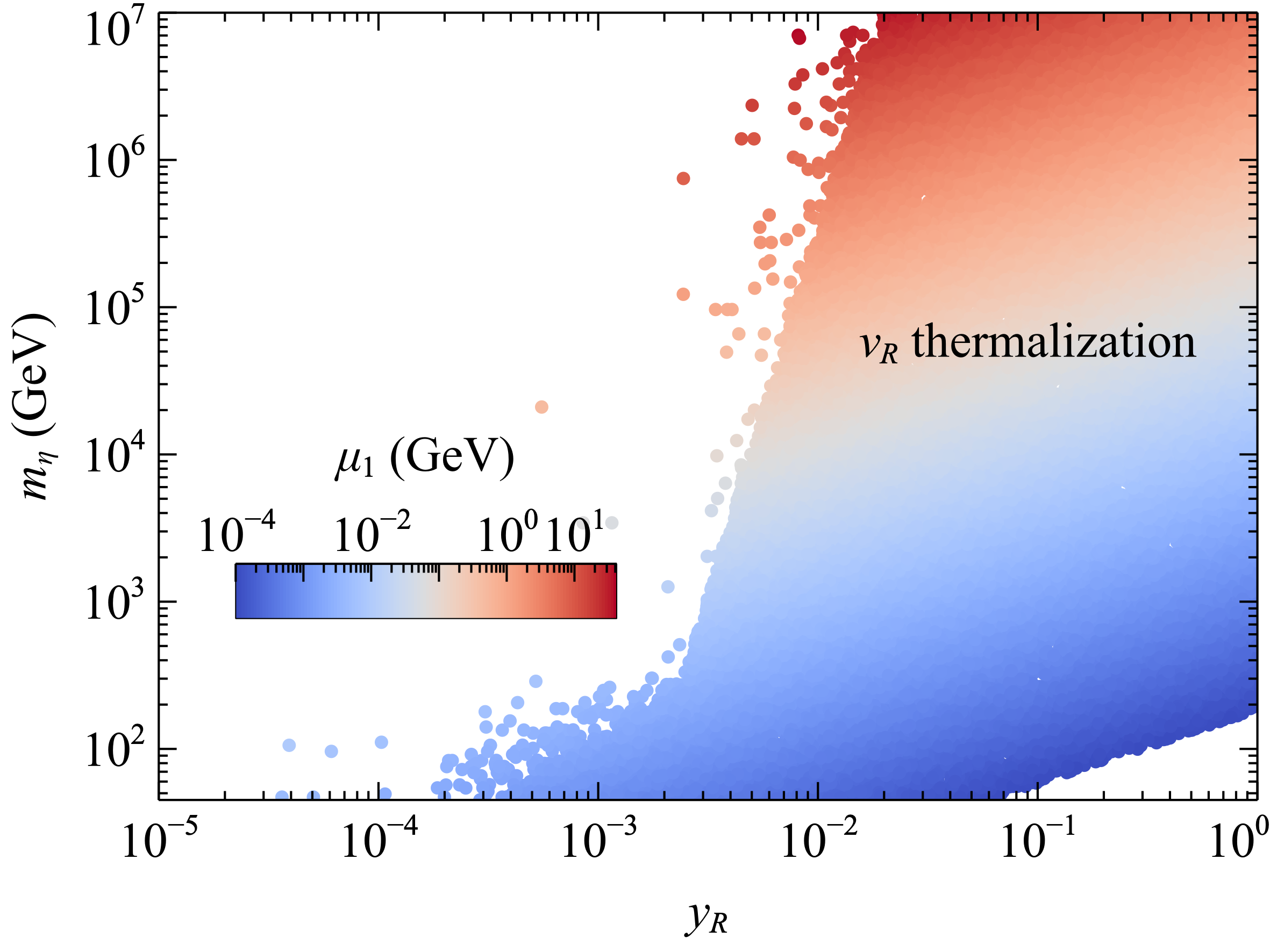} \includegraphics[scale=0.4]{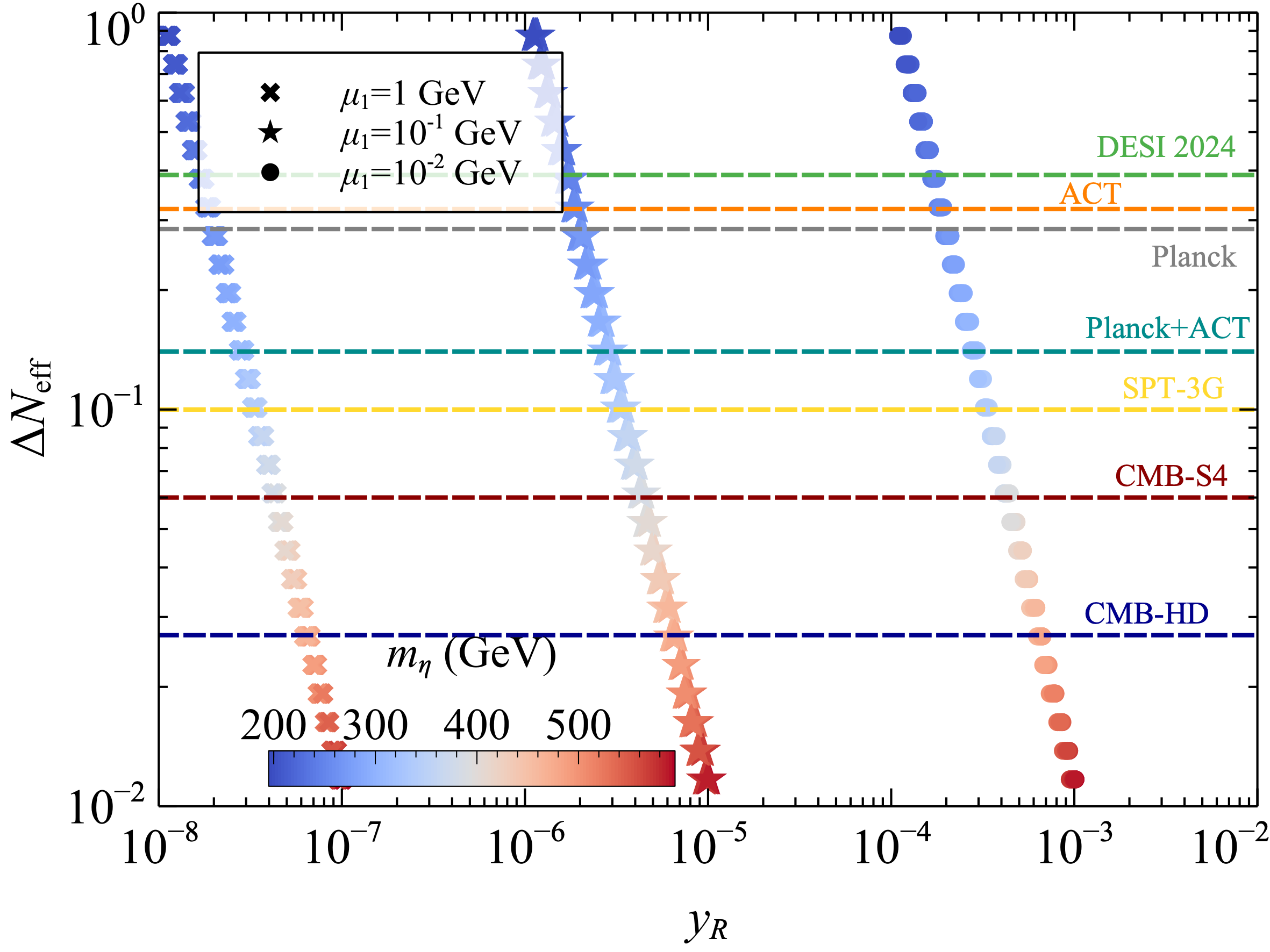}
\caption{[\textit{Left}:] parameter space delineating the thermal and non-thermal regions in the $m_\eta$–$y_R$ plane. The color code denotes $\mu_1$ values. [\textit{Right}:] $\Delta{N}_{\rm eff}$ as a function of $y_R$ for three values of $\mu_1$, as indicated in the inset. Constraints from current CMB observations and projected sensitivities of future CMB experiments are shown by different colored dashed lines.}
\label{fig:neff}
\end{figure*}
In the \textit{left} panel of Fig.~\ref{fig:neff}, we present the region of parameter space in the $m_\eta$–$y_R$ plane where $\nu_R$ becomes thermalized. The color code represents the values of $\mu_1$. Throughout the $\Delta{N_{\rm eff}}$ analysis, we fix the neutrino mass scale to $m_\nu = 0.05$ eV using Eq.~\ref{eq:numass}. The colored region corresponds to parameter values for which $\nu_R$ is thermalized. In this regime, the resulting contribution to $\Delta{N_{\rm eff}}$ is always larger than $\sim 0.135$, and is therefore excluded by combined Planck+ACT \cite{AtacamaCosmologyTelescope:2025nti}. The white region represents the parameter space where a non-thermal contribution to $\Delta{N}_{\rm eff}$ arises from the equilibrium decay of $\eta$. In this case, the evolution of the energy density of $\nu_R$ is governed by the following Boltzmann equation \cite{Biswas:2022vkq,Ma:2025xoj},
\begin{eqnarray}
    \frac{d\rho_{\nu_R}}{dz}=-\frac{4\beta\nu_R}{z}+\frac{1}{z\mathcal{H}(z)}\langle E\Gamma_\eta\rangle n_\eta^{\rm eq},
\end{eqnarray}
where $z=m_\eta/T$, $\beta=1+\frac{T}{3g_{*s}(T)}\frac{dg_{*s}}{dT}$, $\langle E\Gamma_\eta\rangle=2\times2\frac{m_\eta^2y_R^2}{32\pi}$, $\mathcal{H}$ is the Hubble expansion rate, and $n^{\rm eq}_{\eta}$ is the equilibrium number density of $\eta$. The $\Delta{N_{\rm eff}}$ is then computed at CMB epoch as
\begin{eqnarray}
\Delta{N_{\rm eff}}=N_{\nu_R}\frac{\rho_{\nu_R}}{\rho_{\nu_L}}\bigg|_{T_{\rm CMB}},
\end{eqnarray}
where $\rho_{\nu_L}=2\frac{7}{8}\frac{\pi^2}{30}T^4$ is the energy density of the SM neutrinos, and $T_{\rm CMB}\simeq0.26$ eV. In the \textit{right} panel of Fig.~\ref{fig:neff}, we display $\Delta{N_{\rm eff}}$ as a function of $y_R$ for three different choices of $\mu_1$, as indicated in the inset of the figure. The color coding represents the mass of $\eta$. As $y_R$ decreases, the contribution to $\Delta{N}_{\rm eff}$ correspondingly increases. From this analysis, we find that DESI \cite{DESI:2024mwx} excludes the region with $m_\eta<240$ GeV. The Planck \cite{Planck:2018vyg} data exclude $m_{\eta}<260$ GeV. The ACT \cite{AtacamaCosmologyTelescope:2025nti} data exclude $m_{\eta}<250$ GeV, while the combined Planck+ACT \cite{AtacamaCosmologyTelescope:2025nti} analysis excludes $m_{\eta}<300$ GeV. Future measurements by SPT-3G \cite{SPT-3G:2019sok} are expected to exclude $m_{\eta}<330$ GeV, CMB-S4 \cite{Abazajian:2019eic} will probe values down to $m_{\eta}<380$ GeV, and CMB-HD \cite{CMB-HD:2022bsz} will be sensitive to regions with $m_{\eta}<460$ GeV.

%%%%%%%%%%%%%%%%%%%%%%%%%%%%%%%%%%%%%%%%%%%%%%%%%%%%%%%%%%%%%%%%%%%%%%
\section{Freeze-out relic of dark matter}

We assume the reheating temperature of the universe is much larger than the $U(1)_{B-L}$ symmetry-breaking scale, $v_\chi$. As a result, the DM can be thermally populated because of its $B-L$ gauge interactions. Based on the charge of $U(1)_{B-L}$ symmetry breaking scalar ($\chi$), we consider two separate scenarios: A) DM ($S$) is a Dirac fermion ($\chi\sim3$), and B) DM ($S$) is a Majorana fermion ($\chi\sim4$).

\subsection{Dirac dark matter}

We add a singlet vector-like fermion, $S(=S_L+S_R)$, with $B-L$ charge of 2. Since the $B-L$ charge of the scalar $\chi$ is chosen to be 3, it cannot couple to $SS$. Also, the dimension-4 term $\chi \overline{S_L} \nu_R$ is forbidden by $\mathcal{Z}_2$. As a result, $S$ behaves as Dirac DM in this scenario. The DM Lagrangian can be written as
\begin{eqnarray}
    \mathcal{L}^{\rm Dirac}_{\rm DM}=\bar{S}i\gamma^\mu\mathcal{D}_\mu S - m_S\bar{S}S,
\end{eqnarray}
where $\mathcal{D}_\mu=\partial_\mu+i2\textsl{g}_{\rm BL}(Z_{\rm BL})_\mu$. The gauge boson, $Z_{\rm BL}$ acquires a mass, $m_{Z_{\rm BL}}=3\textsl{g}_{\rm BL}v_\chi$ after the breaking of $U(1)_{B-L}$. In this case, the DM $S$ is thermalised through the gauge interactions mediated by $Z_{\rm BL}$. As the temperature falls below the mass scale of $S$, it gets decoupled from the thermal bath, leaving a relic of $S$.

In this scenario, the free parameters relevant for relic density calculation are \{ $m_{S}\equiv m_{\rm DM},v_\chi,\textsl{g}_{\rm BL}$ \}. 
\begin{figure}[h]
    \centering
    \includegraphics[scale=0.4]{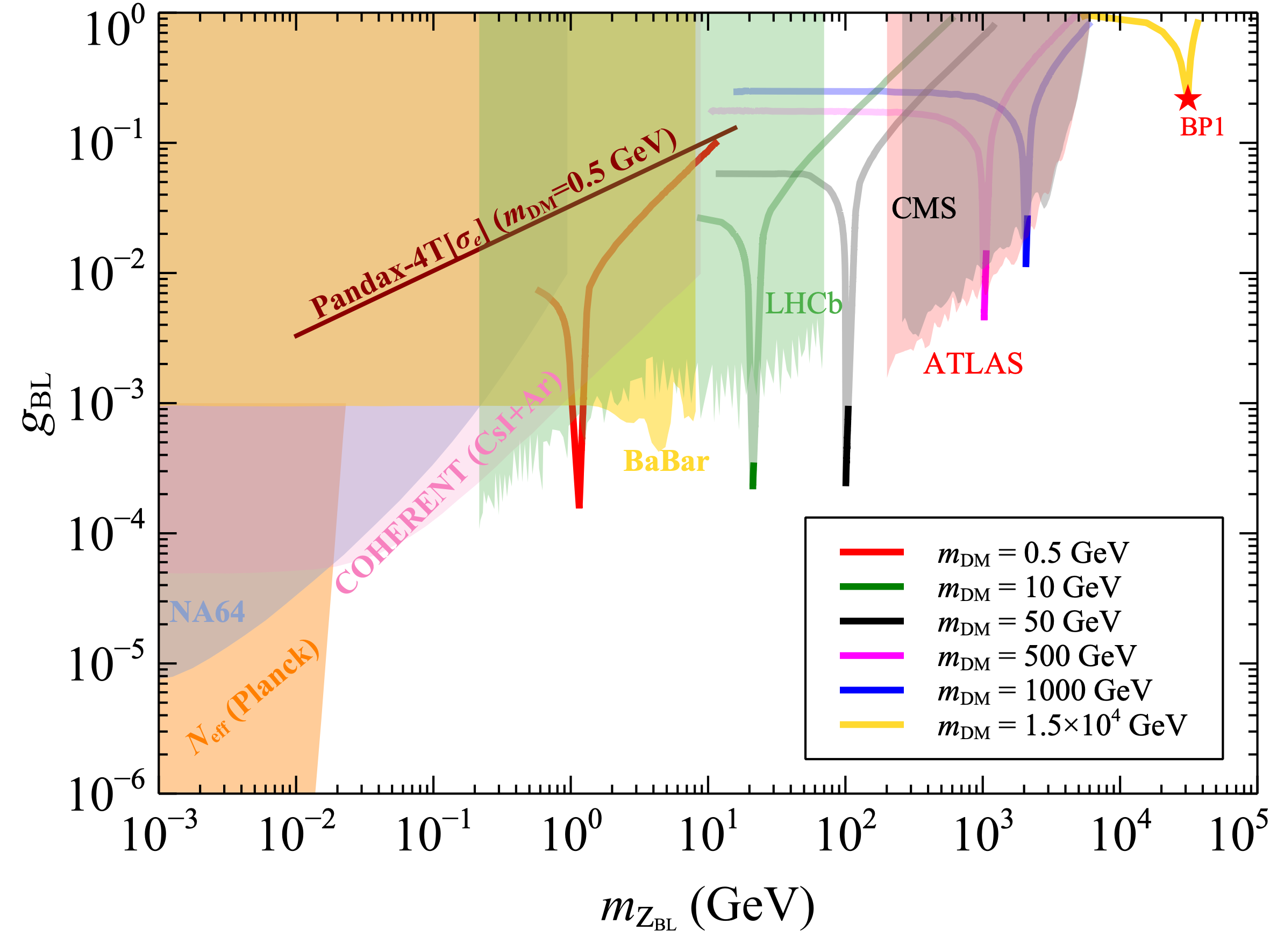}
    \caption{Correct DM relic parameter space in $\textsl{g}_{\rm BL}-m_{Z_{\rm BL}}$ plane for six choices of DM masses shown with different colored lines. Constraints from CMS \cite{CMS:2021ctt}, ATLAS \cite{ATLAS:2019erb}, LHCb \cite{LHCb:2019vmc}, NA64 \cite{Banerjee:2019pds}, BaBar \cite{BaBar:2014zli}, COHERENT \cite{Cadeddu:2020nbr}, $N_{\rm eff}$ (Planck) \cite{Ghosh:2024cxi} are shown with different shaded regions.}
    \label{fig:gblmzbldirac}
\end{figure}
The relic density suffers a sharp fall near the $Z_{\rm BL}$ resonance. The DM relic can be satisfied for a large range of DM mass by utilizing the resonance effect. In the Fig. \ref{fig:gblmzbldirac}, we show the correct relic density parameter space in the plane of $\textsl{g}_{\rm BL}$ vs $m_{Z_{\rm BL}}$ for six different values of DM mass, as shown with different colored lines. The light-colored part of each line is excluded from the direct detection experiment, LZ \cite{LZ:2024zvo}. The dark-colored part of each line, which is the resonance, is allowed by both the direct detection and relic density constraints. For DM mass of 500 MeV, the constraint from the DM-$e$ scattering experiment PandaX-4T \cite{PandaX:2022xqx} is shown with the dark red solid line. The constraint from LHCb \cite{LHCb:2019vmc} is shown with a green shaded region. The CMS \cite{CMS:2021ctt} and ATLAS \cite{ATLAS:2019erb} constraints are shown with gray and red shaded regions, respectively. The constraints from NA64 \cite{Banerjee:2019pds}, BaBar \cite{BaBar:2014zli}, COHERENT \cite{Cadeddu:2020nbr}, and $N_{\rm eff}$ (Planck) \cite{Ghosh:2024cxi} are shown with light blue shaded, yellow shaded, and pink shaded regions, respectively.

%%%%%%%%%%%%%%%%%%%%%%%%%%%%%%%%%%%%%%%%%%%%%%%%%%%%%%%%%%%%%%%%%%%%%%
\subsection{Majorana dark matter}

We now turn to the possibility of a Majorana DM. If we extend the Dirac neutrino mass setup with the choice that $B-L$ charge of $\chi$ is 4 instead of 3, then the introduction of two chiral fermions $S_L$ and $S_R$ having two units of $B-L$ charges results in Majorana fermion DM. These chiral fermions couple to $\chi$ and acquire Majorana masses. They maintain a Dirac mass as well, just as before. On the other hand, $\nu_R$ cannot couple to $\chi$ or $S$, so that neutrinos remain Dirac and $S$ is in the dark sector.
The dark sector Lagrangian can be written as
\begin{eqnarray}
    \mathcal{L}^{\rm Majorana}_{\rm DM}&=&\overline{S_L}i\gamma^\mu\mathcal{D}_\mu S_L+\overline{S_R}i\gamma^\mu\mathcal{D}_\mu S_R- y_1 \overline{S_{L}^C} S_{L} \chi^\dagger\nonumber\\&-&  y_2 \overline{S_{R}^C} S_{R} \chi^\dagger - m_{12} (\overline{S_{R}} S_{L}+\overline{S_{L}} S_{R}) ,
\end{eqnarray}

We can now write the dark sector mass matrix in the basis ($S_L~~S_R^C$) as
\begin{eqnarray}
    \begin{pmatrix}
        \overline{S_L^C}& \overline{S_R}
    \end{pmatrix}.\begin{pmatrix}
        m_L&m_{12}\\
        m_{12}&m_R
    \end{pmatrix}.\begin{pmatrix}
        S_L\\ S^C_R
    \end{pmatrix}
\end{eqnarray}
where $m_L=\sqrt{2}y_1v_\chi, m_R=\sqrt{2}y_2v_\chi$. This mass matrix can be diagonalized with an orthogonal matrix, with a rotation angle given as 
\begin{eqnarray}
    \tan2\theta=\frac{2m_{12}}{m_R-m_L},
\end{eqnarray}
resulting in two Majorana mass states $S_1,S_2$ as $S_i=(S_{iL}+S_{iL}^C)/\sqrt{2}$ where
\begin{eqnarray}
    S_{1L}&=&\cos\theta S_L+\sin\theta S_R^C,\\
    S_{2L}&=&-\sin\theta S_L+\cos\theta S_R^C.
\end{eqnarray}
with masses $m_{S_1}$ and $m_{S_2}$.

The Yukawa couplings $y_1,y_2$ and $m_{12}$ can be expressed in terms of physical masses and mixing angle as
\begin{eqnarray}
    y_1&=&\frac{m_{S_1} \cos^2\theta+ m_{S_2} \sin^2\theta}{\sqrt{2}v_\chi},\\
    y_2&=&\frac{m_{S_2} \cos^2\theta+ m_{S_1} \sin^2\theta}{\sqrt{2}v_\chi},\\
    m_{12}&=&\frac{m_{S_2}- m_{S_1}}{2}\sin2\theta.
\end{eqnarray}

The gauge boson, $Z_{\rm BL}$ acquires a mass, $m_{Z_{\rm BL}}=4\textsl{g}_{\rm BL}v_\chi$ after the breaking of $U(1)_{B-L}$. The free parameters relevant for relic density calculation are \{ $m_{S_1}\equiv m_{\rm DM},v_\chi,m_\chi,\sin\gamma, \textsl{g}_{\rm BL}$ \}.
\begin{figure}[h]
\centering
\includegraphics[scale=0.4]{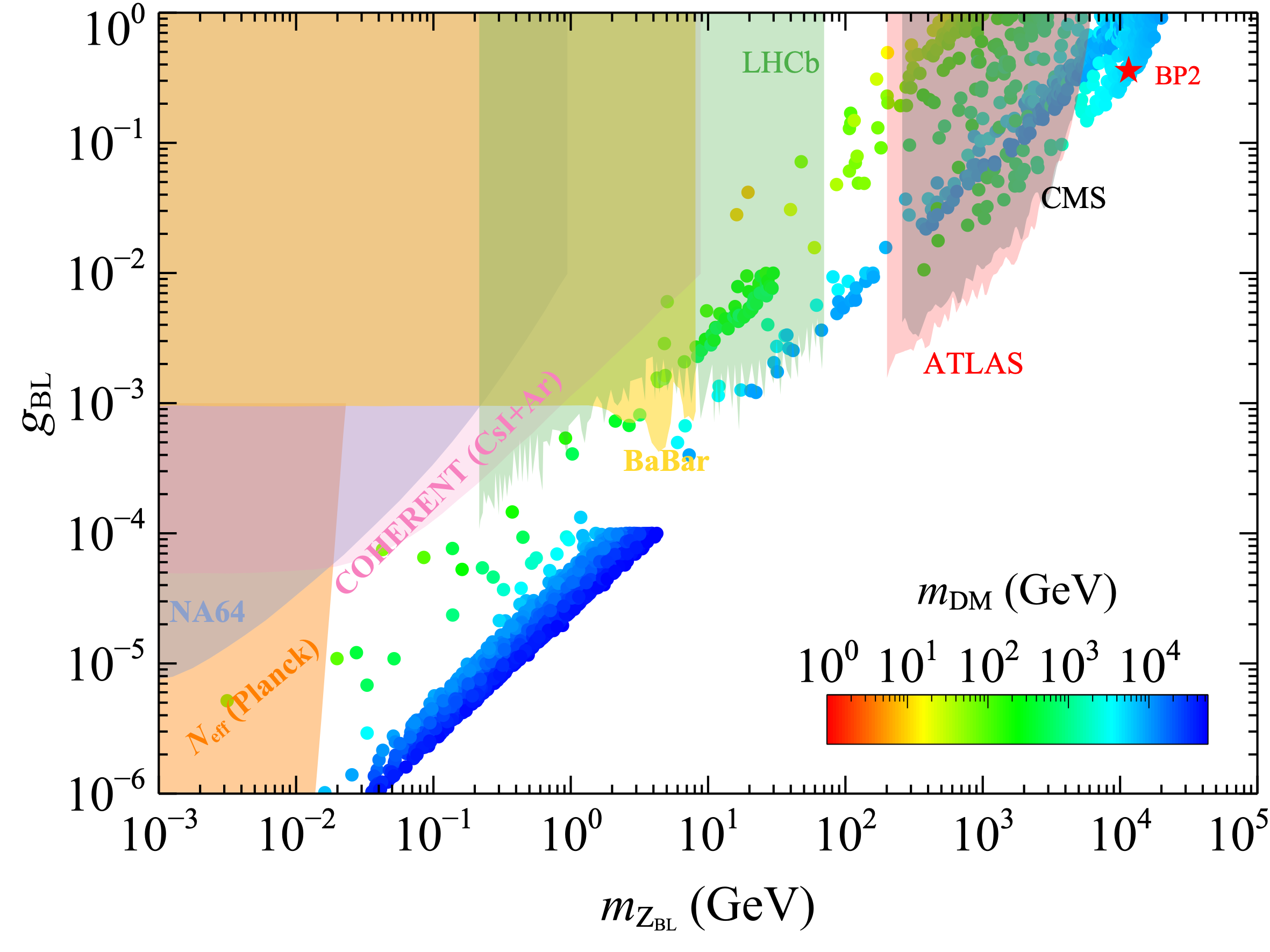}
\caption{Correct DM relic parameter space in $\textsl{g}_{\rm BL}-m_{Z_{\rm BL}}$ plane. Constraints from CMS \cite{CMS:2021ctt}, ATLAS \cite{ATLAS:2019erb}, LHCb \cite{LHCb:2019vmc}, NA64 \cite{Banerjee:2019pds}, BaBar \cite{BaBar:2014zli}, COHERENT \cite{Cadeddu:2020nbr}, $N_{\rm eff}$ (Planck) \cite{Ghosh:2024cxi} are shown with different shaded regions.}
\label{fig:gblmzblMajorana}
\end{figure}
In Fig. \ref{fig:gblmzblMajorana}, we show the points that satisfy both relic density and direct detection constraints in the plane of $\textsl{g}_{\rm BL}$ vs $m_{Z_{\rm BL}}$. The color code denotes the DM mass. Here we have fixed $m_{S_2}-m_{\rm DM}=10$ GeV, $\sin\theta=10^{-2}$. The other parameters are varied as follows: \{ $ m_{\rm DM}\in[1,10^5] $ GeV, $\textsl{g}_{\rm BL}\in[10^{-6},1]$, $\sin\gamma\in[10^{-4},0.7]$, $v_\chi\in[1,10^5]$ GeV \}. In this case, depending on the masses of DM, $\chi$ and $Z_{\rm BL}$ the following channels contribute in the relic density determination: \{$ S_1S_1\rightarrow ff,\chi\chi,\chi h, hh,\chi Z_{\rm BL},h Z_{\rm BL},Z_{\rm BL}  Z_{\rm BL} $\}. As expected, with increasing mass of $Z_{\rm BL}$, the cross-section decreases, which is compensated by increasing $\textsl{g}_{\rm BL}$. The constraint from LHCb \cite{LHCb:2019vmc} is shown as the green shaded region. Constraints from CMS \cite{CMS:2021ctt} and ATLAS \cite{ATLAS:2019erb} are depicted by the gray and red shaded regions, respectively. Bounds from NA64 \cite{Banerjee:2019pds}, BaBar \cite{BaBar:2014zli}, COHERENT \cite{Cadeddu:2020nbr}, and $N_{\rm eff}$ (Planck) \cite{Ghosh:2024cxi} are shown as the light blue, yellow, and pink shaded regions, respectively.
\begin{table*}[htb]
	\centering
	\begin{tabular}{|c|c |c|c|c|c|c|c|c|c|c|c|c|c|} 
		\hline
		BPs & $\textsl{g}_{\rm BL}$ & $v_{\chi}$ (GeV) & $\lambda_{\chi}$ & $m_{\chi}$ (GeV) & $m_{\rm DM}$ (GeV) & $T_c$ (GeV) & $T_n$ (GeV) & $T_p$ (GeV) & $T_{\rm reh}$ (GeV) & $\alpha_p$ & $\beta/\mathcal{H}_p$  \\ 
		\hline
		BP1  & $0.217767$ & $47756$ & $0.0035536$& $4026.09$ & $1.5\times10^4$ & $9793.52$ &
		$2134.87$ & 2024.88 & 3043.04 & $0.8063$& $288.67$    \\
		\hline
		BP2  & 0.358904 & 8042.41 & 0.071633 & 3044.09  & 5719.44 & 2915.57  & 1907.68& 1820.76 & 1854.05
		& 0.0865 & 253.15     \\
		\hline
	\end{tabular}
	\caption{Benchmark points giving rise to observable gravitational wave signatures while satisfying constraints from DM relic density, direct detection, and colliders. The BP1 and BP2 are shown with red stars in Fig. \ref{fig:gblmzbldirac} and \ref{fig:gblmzblMajorana}, respectively.}\label{tab:tab2}
\end{table*}

%%%%%%%%%%%%%%%%%%%%%%%%%%%%%%%%%%%%%%%%%%%%%%%%%%%%%%%%%%%%%%%%%%%%%%
\section{Gravitational waves from $U(1)_{B-L}$ symmetry breaking}

The model also presents intriguing cosmological detection prospects through stochastic gravitational waves, providing a complementary probe of its viability. These gravitational wave signatures originate from the phase transition associated with the spontaneous breaking of $U(1)_{B-L}$. This symmetry is broken when the scalar $\chi$ acquires a vev. A first-order phase transition (FOPT) can occur if the true vacuum, where $U(1)_{B-L}$ is broken, has a lower energy density than the high-temperature false vacuum, with a potential barrier separating them. To get the parameter space in which an FOPT occurs, we analyze the structure of the effective potential incorporating the tree-level potential $V_{\rm tree}$, the one-loop Coleman-Weinberg correction $V_{\rm CW}$ \cite{Coleman:1973jx}, and finite-temperature corrections~\cite{Dolan:1973qd,Quiros:1999jp}.

The critical temperature $T_c$, at which the potential develops two degenerate minima $(0, v_c)$, is determined by studying the temperature evolution of the potential. The ratio $v_c/T_c$ serves as the order parameter, with larger values indicating a stronger first-order phase transition (FOPT). The FOPT proceeds via quantum tunneling, with the tunneling rate estimated by calculating the bounce action $S_3$. The nucleation temperature $T_n$ is then obtained by equating the tunneling rate per unit volume to the Hubble expansion rate of the universe, $\Gamma (T_n) = \mathcal{H}^4(T_n)$.

For a strongly supercooled transition (such as BP1),
characterized by $T_n/T_c \ll 1$ and $\alpha_n \gg 1$\footnote{$\alpha_n$ is the strength of the FOPT computed at the nucleation temperature, $\alpha_n=\epsilon(T_n)/\rho_{\rm rad}(T_n)$, $\epsilon$ is the latent heat. For the BP1, $\alpha_n=3.2059$ and for BP2 it is computed to be 0.0741.}, the standard
treatment evaluated at $T_n$ is insufficient. The phase transition does not complete at $T_n$, it completes at the percolation temperature $T_p < T_n$, defined as the temperature at which enough bubbles of true vacuum have nucleated and grown to fill $\sim 71\%$ of space. Furthermore, when the bubble network percolates at $T_p$, the latent heat is released into the plasma, reheating the universe to the temperature $T_{\rm reh} > T_p$. Since gravitational waves are produced at bubble collision and subsequently redshift from $T_{\rm reh}$, the GW spectrum must be evaluated using the parameters: strength of the transition ($\alpha_p$) and inverse duration of the transition ($\beta/\mathcal{H}|_{T_p}$) defined at $T_p$, with $T_{\rm reh}$ as the reference temperature. The details of this analysis are given in Appendix \ref{app:supercool}

We then compute the key parameters required to estimate the stochastic gravitational wave (GW) spectrum originating from bubble collisions~\cite{Turner:1990rc,Kosowsky:1991ua,Kosowsky:1992rz,Kosowsky:1992vn,Turner:1992tz}, sound waves in the plasma~\cite{Hindmarsh:2013xza,Giblin:2014qia,Hindmarsh:2015qta,Hindmarsh:2017gnf}, and plasma turbulence~\cite{Kamionkowski:1993fg,Kosowsky:2001xp,Caprini:2006jb,Gogoberidze:2007an,Caprini:2009yp,Niksa:2018ofa}. The stochastic GW energy density receives contributions from three main sources: bubble wall collisions, sound waves in the plasma, and magnetohydrodynamic turbulence. The total GW spectrum can be expressed as the sum of these individual components:
\begin{eqnarray}
\Omega_{\rm GW}h^2 \approx \Omega_{\rm col} h^2 + \Omega_{\rm sw} h^2 + \Omega_{\rm turb} h^2.
\end{eqnarray}
\begin{figure}[h]
    \centering
    \includegraphics[scale=0.4]{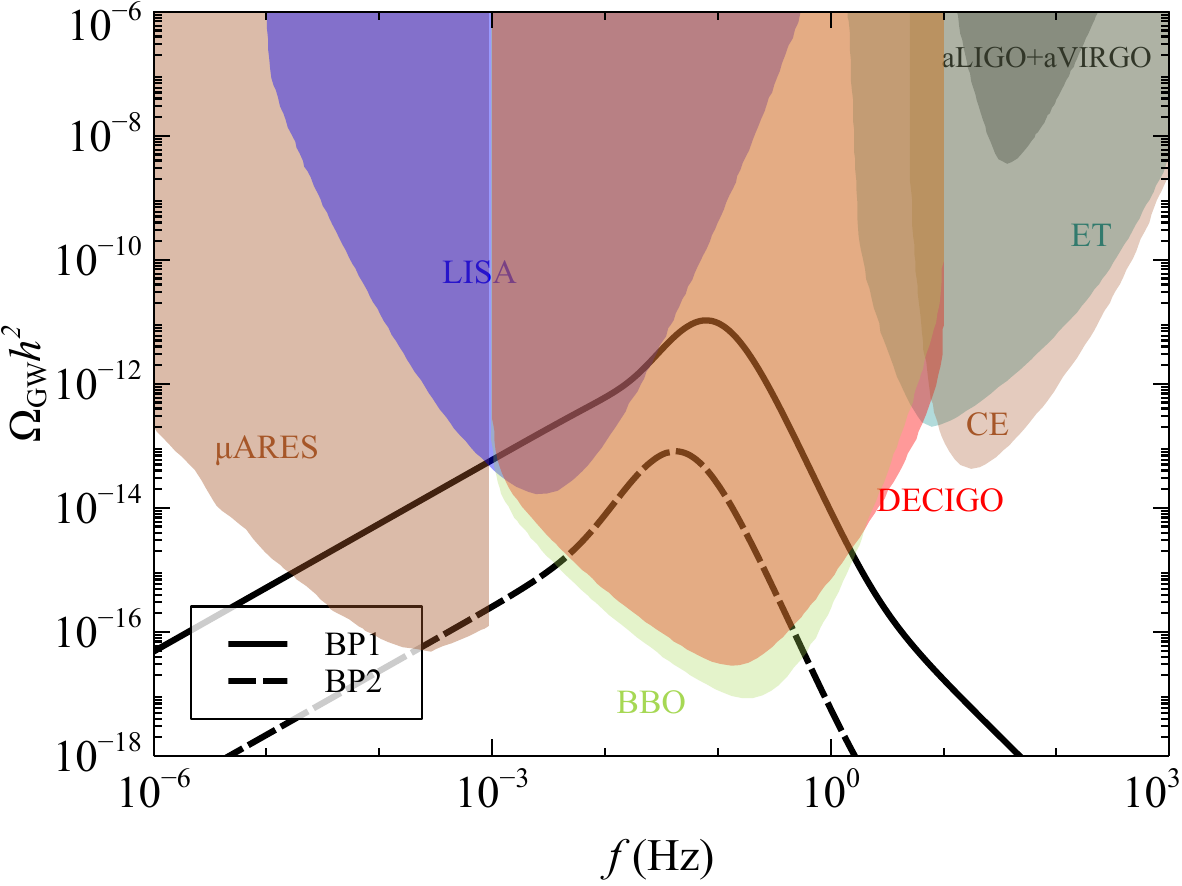}
    \caption{Gravitational wave spectrum for benchmark points satisfying DM relic as given in Table \ref{tab:tab2}.}
    \label{fig:gw}
\end{figure}
We choose two benchmark points for GW from the Dirac DM (BP1) and Majorana DM (BP2) scenarios discussed above. In case of Dirac DM, as there is no interaction between the scalar field $\chi$ and the fermion, $S$, the FOPT is driven by the parameters \{$v_\chi,\textsl{g}_{\rm BL}$, $\lambda_\chi$\} only. On the other hand, in the case of Majorana DM, the Majorana fermions interact with $\chi$ and modify the Coleman-Weinberg potential. The FOPT sensitive parameters in this case are \{$v_\chi,\textsl{g}_{\rm BL},\lambda_\chi, y_1,y_2$\}. The parameters are given in Table \ref{tab:tab2}. The values of $y_1,y_2$ corresponding to BP2 are 0.502866 and 0.503746, respectively. 

BP1 is found to be strongly supercooled, with $T_n/T_c = 0.22$ and $\alpha_n = 3.21$, indicating that the vacuum energy dominates over the radiation energy density already at $T_n$. In this regime, the standard GW analysis evaluated at $T_n$ is insufficient, and a careful percolation analysis is required. We compute the percolation temperature $T_p$ using the full Hubble rate including the vacuum energy contribution as detailed in Appendix \ref{app:supercool}. This yields $T_p = 2024.88$~GeV and, from energy conservation, the reheating temperature $T_{\rm reh} = 3043.04$~GeV. The strength parameter $\alpha_p = 0.8063 < 1$ confirms that at $T_p$ the transition is no longer vacuum dominated, percolation completes successfully. For BP2, with $T_n/T_c = 0.65$ and $\alpha_n = 0.074$, the transition is only mildly supercooled and the corrections are moderate: $T_p = 1820.76$~GeV, $T_{\rm reh} =
1854.05$~GeV, and $\alpha_p = 0.087$.

In Fig. \ref{fig:gw}, we show the gravitational wave amplitude as a function of frequency for BP1 and BP2 as mentioned in Table \ref{tab:tab2}. Sensitivities from LISA \cite{LISA:2017pwj}, DECIGO \cite{Seto:2001qf}, $\mu$ARES \cite{Sesana:2019vho}, BBO \cite{Corbin:2005ny},  CE \cite{LIGOScientific:2016wof}, ET \cite{Punturo:2010zz}, aLIGO, aVIRGO \cite{LIGOScientific:2016wof} are shown with different colored shaded regions.
The peak amplitude lies in the sensitivity ranges of LISA, DECIGO, and BBO.

%%%%%%%%%%%%%%%%%%%%%%%%%%%%%%%%%%%%%%%%%%%%%%%%%%%%%%%%%%%%%%%%%%%%%%
\section{Conclusions}

In this paper, we studied a variant of gauged ${B-L}$ symmetric model where the singlet scalar $\chi$, responsible for breaking the $U(1)_{B-L}$ symmetry, has $B-L$ charge 3 or 4. As a result, the right-handed neutrinos ($\nu_R$ with -1 $B-L$ charge) do not acquire any Majorana masses even after the $B-L$ symmetry is broken. If the $B-L$ charge of $\chi$ is 3, then we find a possibility of a Dirac fermionic dark matter. On the other hand, if the $B-L$ charge of $\chi$ is 4, then the DM can have both Dirac as well as Majorana masses, leading to a pseudo-Dirac DM candidate. In either case, the correct relic density of DM can be achieved through freeze-out of various processes. The model further predicts distinctive cosmological signatures in the form of stochastic gravitational waves generated during the first-order phase transition associated with the breaking of the $U(1)_{B-L}$ symmetry, which can be probed at present and future gravitational wave experiments. In addition, the presence of light right-handed neutrinos in this framework leads to a potentially observable contribution to $\Delta N_{\rm eff}$ through their thermal or non-thermal production in the early Universe. Current and future CMB observations therefore provide a complementary and powerful probe of this scenario, offering an independent test of the Dirac neutrino sector beyond laboratory experiments.

%%%%%%%%%%%%%%%%%%%%%%%%%%%%%%%%%%%%%%%%%%%%%%%%%%%%%%%%%%%%
\section*{Acknowledgment}
P.K.P. acknowledges the Ministry of Education, Government of India, for providing financial support for his research via the Prime Minister’s Research Fellowship (PMRF) scheme.


%%%%%%%%%%%%%%%%%%%%%%%%%%%%%%%%%%%%%%%%%%%%%%%%%%%%%%%%%%%%%%%%%%%%%%

\appendix
\section{Feynman diagrams responsible for thermalization of $\nu_R$}
\begin{figure}[h]
    \centering
    \includegraphics[scale=0.5]{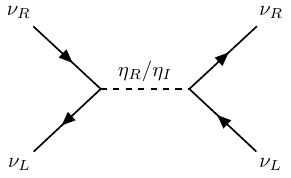}
    \includegraphics[scale=0.5]{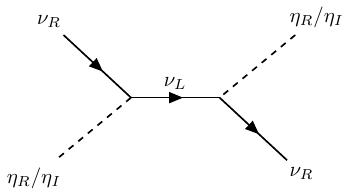}
    \includegraphics[scale=0.5]{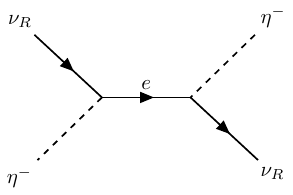}
    \includegraphics[scale=0.6]{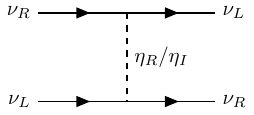}
    \includegraphics[scale=0.6]{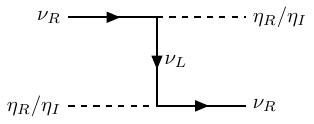}
    \includegraphics[scale=0.6]{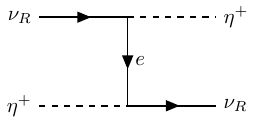}
    
    \caption{Diagrams contributing to the thermalization of $\nu_R$.}
    \label{fig:nuRequilibirum}
\end{figure}

In Fig.~\ref{fig:nuRequilibirum}, we present the Feynman diagrams of the processes responsible for maintaining thermal equilibrium among $\nu_R$ and the Standard Model bath.

\section{Details of supercooled first-order phase transition}\label{app:supercool}

The spontaneous breaking of the $U(1)_{B-L}$ symmetry occurs through a first-order phase transition once the scalar $\chi$ develops a vacuum expectation value. In the usual approach, the resulting gravitational-wave spectrum is computed at the nucleation temperature $T_n$, defined as the temperature at which, on average, one bubble nucleates within one Hubble volume during one Hubble time. This treatment, however, becomes unreliable in the strongly supercooled regime, where $T_n/T_c \ll 1$ and $\alpha_n \gg 1$. In such cases, a more refined description of the phase transition is necessary. In this appendix, we briefly describe the improved treatment based on the percolation temperature $T_p$ and the reheating temperature $T_{\rm reh}$, and apply it to both benchmark points in Table \ref{tab:tab2}.
The thermal nucleation rate of true-vacuum bubbles per unit volume is
given by
\begin{eqnarray}
    \Gamma=T^4 e^{-S_3/T}\left(\frac{S_3/T}{2\pi}\right)^{3/2}
\end{eqnarray}
where $S_3$ is the bounce action. The nucleation temperature $T_n$ is then defined by the condition $\Gamma(T_n)/\mathcal{H}(T_n)^4 \sim 1$. While $T_n$ marks the onset of nucleation, it does not describe when the transition actually completes. To quantify this, we compute the average number of bubbles nucleated per unit volume, weighted by their physical volume \cite{Guth:1979bh,Guth:1981uk,Ellis:2019oqb}
\begin{eqnarray}
    I(T)=\frac{4\pi v_w^3}{3}\int_T^{T_c}\frac{dT^\prime}{T^{\prime4}\mathcal{H}(T^\prime)}\Gamma(T^\prime)\left( \int_T^{T^\prime}\frac{dT^{\prime\prime}}{\mathcal{H}(T^{\prime\prime})} \right)^3,
\end{eqnarray}
where $v_w \simeq 1$ is the bubble wall velocity and the Hubble rate $\mathcal{H}$ includes
the contribution from the false vacuum energy $\Delta V(T) =
V(0,T)-V(v_\chi,T)$ along with the radiation energy density $\rho_{\rm rad}$, which can dominate over the radiation energy
density in the supercooled regime and is given as
\begin{eqnarray}
    \mathcal{H}=\sqrt{\frac{\rho_{\rm rad}(T)+\Delta V(T)}{3M_{\rm pl}^2}},
\end{eqnarray}
with $M_{\rm pl}=2.435\times10^{18}$ GeV is the reduced Planck mass. The percolation temperature $T_p$ is defined as the temperature at
which enough bubbles have nucleated and expanded to fill approximately
$71\%$ of space, so that the bubble network percolates and the
transition completes throughout the universe. This condition is
expressed in terms of the false vacuum fraction as
\begin{eqnarray}
    P_{\rm false}(T_p)=e^{-I(T_p)}\approx0.71=e^{-0.34}
\end{eqnarray}
Once percolation completes at $T_p$, the latent heat stored in the
false vacuum, $\Delta V(T_p)$, is rapidly converted into radiation
through bubble collisions. This reheats the plasma from $T_p$ to
the reheating temperature $T_{\rm reh}$, determined by energy
conservation,
\begin{eqnarray}
    \rho_{\rm rad}(T_{\rm reh})=\rho_{\rm rad}(T_p)+\Delta V(T_p)
\end{eqnarray}
Since gravitational waves are produced at the moment of bubble
collision and subsequently redshift from $T_{\rm reh}$ when the
universe re-enters radiation domination, the GW spectrum must be
evaluated using parameters defined at $T_p$ and $T_{\rm reh}$
rather than $T_n$. The phase transition strength parameter is
\begin{eqnarray}
\alpha_p=\frac{\epsilon(T_p)}{\rho_{\rm rad} (T_{\rm reh})},
\end{eqnarray}
where latent heat $\epsilon$ is given as
\begin{eqnarray}
    \epsilon(T_p)=\Delta V(T_p)-\frac{T_p}{4}\frac{d\Delta V}{dT}\Bigg|_{T_p},
\end{eqnarray}
and the inverse duration of the transition is
\begin{eqnarray}
    \frac{\beta}{\mathcal{H}}\Bigg|_{T_p}=T_p\frac{d(S_3/T)}{dT}\Bigg|_{T_p}.
\end{eqnarray}
\begin{figure*}[h]
\centering
\includegraphics[scale=0.4]{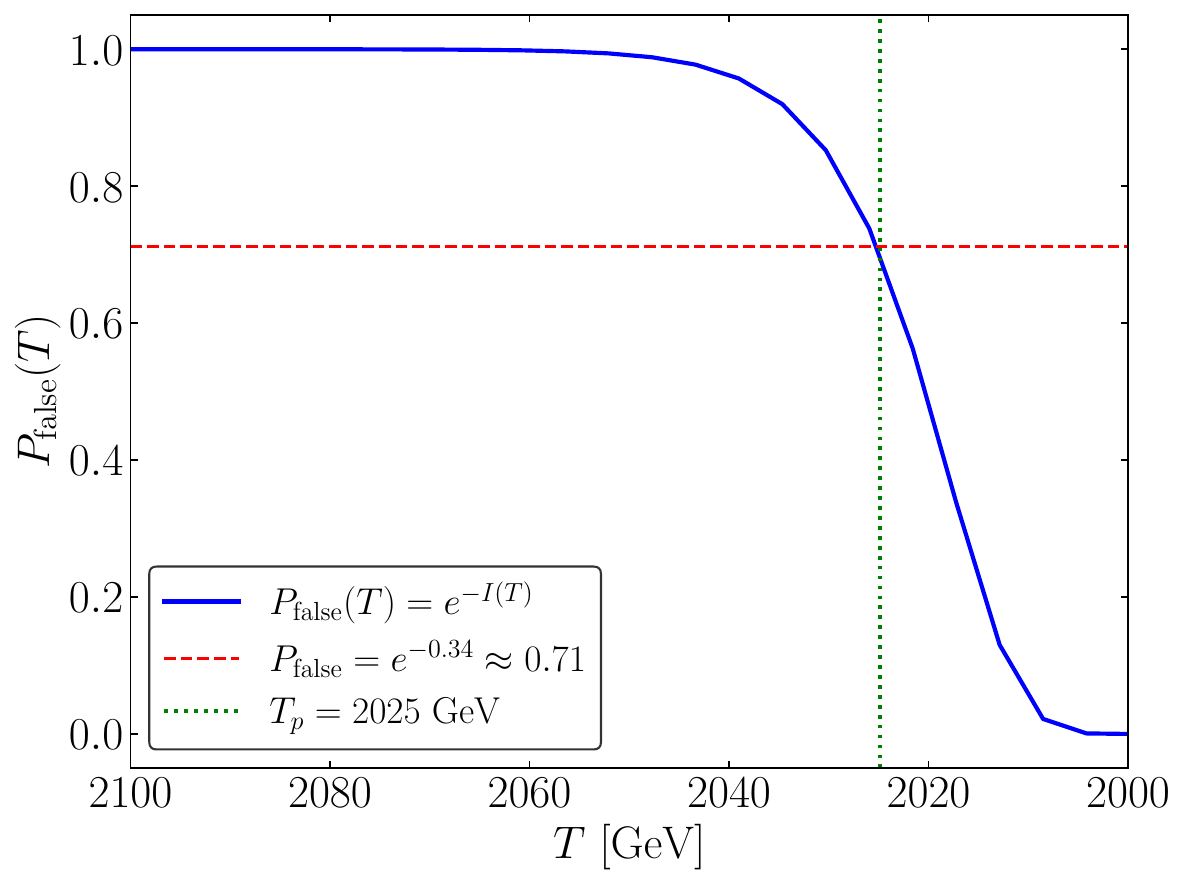}
\includegraphics[scale=0.4]{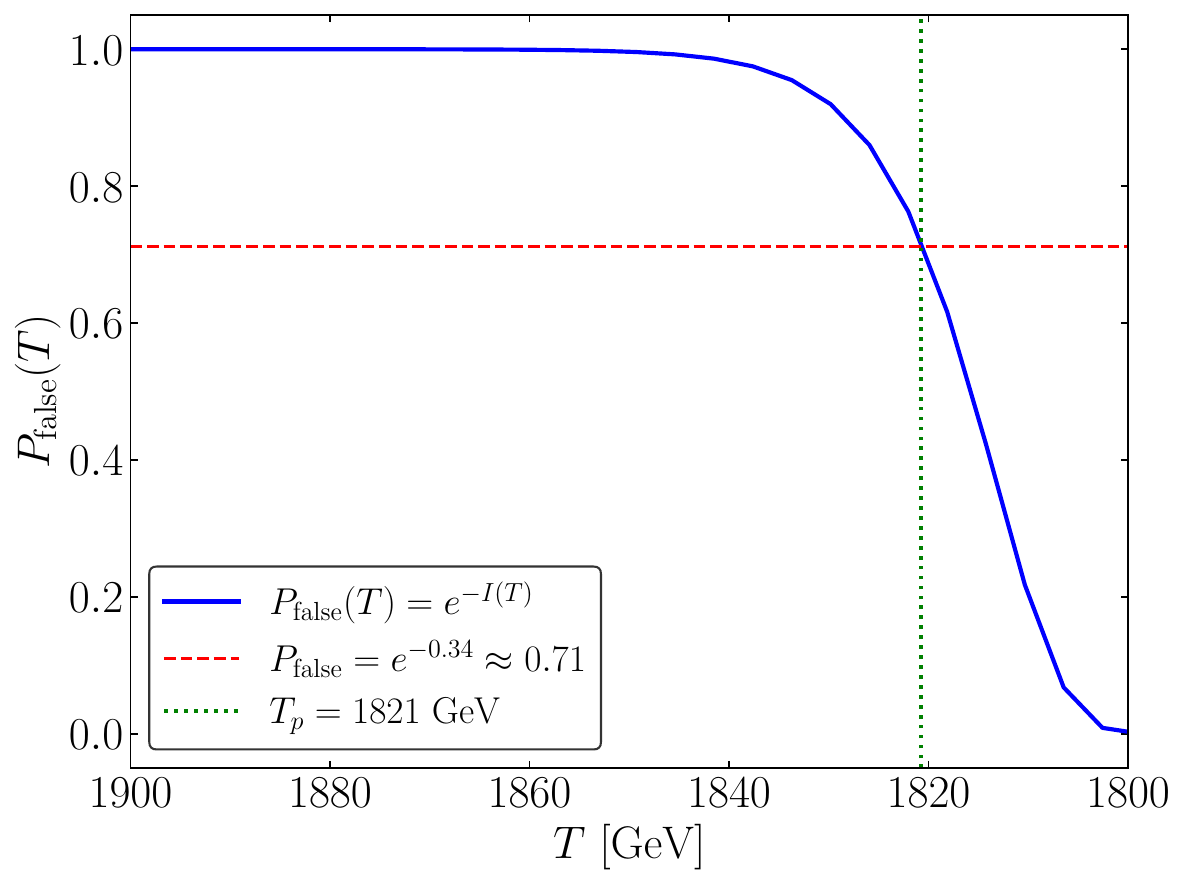}
\caption{False vacuum fraction $P_{\rm false}(T) = e^{-I(T)}$ as a function of temperature for BP1 (left) and BP2 (right). The red dashed line marks the percolation threshold $P_{\rm false} = e^{-0.34} \approx 0.71$. The vertical green dotted line indicates $T_p$.}
\label{fig:perco}
\end{figure*}
In Fig.~\ref{fig:perco}, we show $P_{\rm false}(T)$ as a function of temperature for both benchmark points. The red dashed line marks the percolation threshold $P_{\rm false} = e^{-0.34} \approx 0.71$. The vertical green dotted line indicates $T_p$.

\end{document}